\documentclass[12pt]{article}
\usepackage{amsfonts}
\usepackage{amssymb}
\usepackage{graphics,psboxit,amsmath,amscd}
\usepackage{subfigure}
\usepackage{graphicx}

\def\hybrid{\topmargin -20pt    \oddsidemargin 0pt
        \headheight 0pt \headsep 0pt
        \textwidth 6.35in       
        \textheight 9.25in       
        \marginparwidth .875in
        \parskip 5pt plus 1pt   \jot = 1.5ex}

\hybrid

\def\baselinestretch{1.2}

\catcode`\@=11

\def\marginnote#1{}
\newcount\hour
\newcount\minute
\newtoks\amorpm
\hour=\time\divide\hour by60
\minute=\time{\multiply\hour by60 \global\advance\minute by-\hour}
\edef\standardtime{{\ifnum\hour<12 \global\amorpm={am}%
        \else\global\amorpm={pm}\advance\hour by-12 \fi
        \ifnum\hour=0 \hour=12 \fi
        \number\hour:\ifnum\minute<10 0\fi\number\minute\the\amorpm}}
\edef\militarytime{\number\hour:\ifnum\minute<10 0\fi\number\minute}

\def\draftlabel#1{{\@bsphack\if@filesw {\let\thepage\relax
   \xdef\@gtempa{\write\@auxout{\string
      \newlabel{#1}{{\@currentlabel}{\thepage}}}}}\@gtempa
   \if@nobreak \ifvmode\nobreak\fi\fi\fi\@esphack}
        \gdef\@eqnlabel{#1}}
\def\@eqnlabel{}
\def\@vacuum{}
\def\draftmarginnote#1{\marginpar{\raggedright\scriptsize\tt#1}}

\def\draft{\oddsidemargin -.5truein
        \def\@oddfoot{\sl preliminary draft \hfil
        \rm\thepage\hfil\sl\today\quad\militarytime}
        \let\@evenfoot\@oddfoot \overfullrule 3pt
        \let\label=\draftlabel
        \let\marginnote=\draftmarginnote
   \def\@eqnnum{(\theequation)\rlap{\kern\marginparsep\tt\@eqnlabel}%
\global\let\@eqnlabel\@vacuum}  }

\def\preprint{\twocolumn\sloppy\flushbottom\parindent 2em
        \leftmargini 2em\leftmarginv .5em\leftmarginvi .5em
        \oddsidemargin -.5in    \evensidemargin -.5in
        \columnsep .4in \footheight 0pt
        \textwidth 10.in        \topmargin  -.4in
        \headheight 12pt \topskip .4in
        \textheight 6.9in \footskip 0pt
        \def\@oddhead{\thepage\hfil\addtocounter{page}{1}\thepage}
        \let\@evenhead\@oddhead \def\@oddfoot{} \def\@evenfoot{} }

\def\numberbysection{\@addtoreset{equation}{section}
        \def\theequation{\thesection.\arabic{equation}}}

\def\underline#1{\relax\ifmmode\@@underline#1\else
        $\@@underline{\hbox{#1}}$\relax\fi}

\def\titlepage{\@restonecolfalse\if@twocolumn\@restonecoltrue\onecolumn
     \else \newpage \fi \thispagestyle{empty}\c@page\z@
        \def\thefootnote{\fnsymbol{footnote}} }

\def\endtitlepage{\if@restonecol\twocolumn \else \newpage \fi
        \def\thefootnote{\arabic{footnote}}
        \setcounter{footnote}{0}}  

\catcode`@=12
\relax

\def\figcap{\section*{Figure Captions\markboth
        {FIGURECAPTIONS}{FIGURECAPTIONS}}\list
        {Figure \arabic{enumi}:\hfill}{\settowidth\labelwidth{Figure
999:}
        \leftmargin\labelwidth
        \advance\leftmargin\labelsep\usecounter{enumi}}}
 \relax
\def\tablecap{\section*{Table Captions\markboth
        {TABLECAPTIONS}{TABLECAPTIONS}}\list
        {Table \arabic{enumi}:\hfill}{\settowidth\labelwidth{Table
999:}
        \leftmargin\labelwidth
        \advance\leftmargin\labelsep\usecounter{enumi}}}
 \relax
\def\reflist{\section*{References\markboth
        {REFLIST}{REFLIST}}\list
        {[\arabic{enumi}]\hfill}{\settowidth\labelwidth{[999]}
        \leftmargin\labelwidth
        \advance\leftmargin\labelsep\usecounter{enumi}}}
 \relax
\makeatletter
\newcounter{pubctr}
\def\publist{\@ifnextchar[{\@publist}{\@@publist}}
\def\@publist[#1]{\list
        {[\arabic{pubctr}]\hfill}{\settowidth\labelwidth{[999]}
        \leftmargin\labelwidth
        \advance\leftmargin\labelsep
        \@nmbrlisttrue\def\@listctr{pubctr}
        \setcounter{pubctr}{#1}\addtocounter{pubctr}{-1}}}
\def\@@publist{\list
        {[\arabic{pubctr}]\hfill}{\settowidth\labelwidth{[999]}
        \leftmargin\labelwidth
        \advance\leftmargin\labelsep
        \@nmbrlisttrue\def\@listctr{pubctr}}}
 \relax
\makeatother
\newskip\humongous \humongous=0pt plus 1000pt minus 1000pt

\newif\ifdtup

\relax

\def\be{\begin{equation}}
\def\ee{\end{equation}}
\def\ba{\begin{eqnarray}}
\def\ea{\end{eqnarray}}

\def\a{\alpha}

\def\b{\beta}

\def\g{\gamma}

\def\e{\epsilon}

\def\th{\theta}

\def\l{\lambda}

\def\s{\sigma}

\def\no{\noindent}

\def\qq{\qquad}

\def\IR{\relax{\rm I\kern-.18em R}}
\def\II{\relax{\rm 1\kern-.35em1}}

\renewcommand{\theequation}{\thesection.\arabic{equation}}
\csname @addtoreset\endcsname{equation}{section}

\def\IR{\relax{\rm I\kern-.18em R}}
\def\inv{^{\raise.15ex\hbox{${\scriptscriptstyle -}$}\kern-.05em 1}}

\begin{document}

\begin{titlepage}
\begin{center}

\hfill CERN-PH-TH/2006-140\\
\vskip -.1 cm
\hfill IFT-UAM/CSIC-06-40\\
\vskip -.1 cm
\hfill hep--th/0608029\\

\vskip .5in

{\LARGE The magnon kinematics of the AdS/CFT correspondence}
\vskip 0.4in

{\bf C\'esar G\'omez}$^{1,2}$\phantom{x}and\phantom{x}{\bf Rafael Hern\'andez}$^2$
\vskip 0.1in

${}^1\!$
Instituto de F\'{\i}sica Te\'orica UAM/CSIC\\
Universidad Aut\'onoma de Madrid,
Cantoblanco, 28049 Madrid, Spain\\
{\footnotesize{\tt cesar.gomez@uam.es}}

\vskip .2in

${}^2\!$
Theory Group, Physics Department, CERN\\
CH-1211 Geneva 23, Switzerland\\
{\footnotesize{\tt rafael.hernandez@cern.ch}}

\end{center}

\vskip .4in

\centerline{\bf Abstract}
\vskip .1in
\no
The planar dilatation operator of ${\cal N}=4$ supersymmetric Yang-Mills is the 
hamiltonian of an integrable spin chain whose length is allowed to fluctuate. We 
will identify the dynamics of length fluctuations of planar ${\cal N}=4$ Yang-Mills 
with the existence of an abelian Hopf algebra ${\cal Z}$ symmetry with non-trivial 
co-multiplication and antipode. The intertwiner conditions for this Hopf algebra will 
restrict the allowed magnon irreps to those leading to the magnon dispersion relation. 
We will discuss magnon kinematics and crossing symmetry on the spectrum of ${\cal Z}$. 
We also consider general features of the underlying Hopf algebra with ${\cal Z}$ as 
central Hopf subalgebra, and discuss the giant magnon semiclassical regime. 
\noindent


\end{titlepage}

\def\baselinestretch{1.2}


\baselineskip 20pt


\section{Introduction}

The appearance of integrable structures on both sides of the AdS/CFT correspondence 
has played a central role in our current understanding of the duality. The dilatation 
operator of planar ${\cal N}=4$ supersymmetric Yang-Mills has been shown to correspond, 
at one-loop for the complete theory \cite{MZ}-\cite{psu}, and at several loops in some 
subsectors \cite{dilatation}-\cite{Serban}, to the hamiltonian of a one-dimensional 
integrable system. On the string theory side integrability of the classical sigma model 
on $AdS_5 \times S^5$ \cite{Polchinski} allowed a resolution of the theory in terms of 
spectral curves \cite{Kazakov}-\cite{curves}. The integral equations satisfied by the 
spectral density suggested soon after a discrete Bethe ansatz for the quantum string 
sigma model \cite{qBethe}. The Bethe equations of the gauge theory were then shown to 
arise from an asymptotic $S$-matrix in \cite{StaudacherS}, and the $S$-matrix of 
${\cal N}=4$ Yang-Mills was recently derived in \cite{BeisertS}. Integrability is 
thus encoded in a factorizable $S$-matrix on both sides of the correspondence. 
The string theory $S$-matrix describes the scattering of some classical 
lumps supported on the two-dimensional worldsheet. A semiclassical description at 
strong 't~Hooft coupling of this $S$-matrix has been proposed in \cite{magnons} 
(see also \cite{Dorey}-\cite{Kruczenski}), based on the classical equivalence of 
strings moving on $\mathbb{R} \times S^2$ and the sine-Gordon integrable model 
\cite{Pohlmeyer}, \cite{Mikhailov}. On the gauge theory side the $S$-matrix describing 
the scattering of magnon fluctuations of the spin chain can be constrained by the 
symmetries of the system and by the Yang-Baxter triangular equation \cite{BeisertS}. 
However these symmetries are not enough to fix completely the scattering matrix, and 
additional physical requirements such as unitarity, bootstrap in the case of a 
non-trivial spectrum of bound states, and crossing symmetry, need to be 
imposed \cite{ZZ}.
  
The $S$-matrix for the quantum string 
Bethe ansatz should extrapolate at weak but finite coupling to the gauge theory 
$S$-matrix through a dressing phase factor \cite{qBethe}, 
\be
S_{\hbox{\tiny{string}}}(p_j,p_k) = e^{i \, \theta(p_j,p_k)} \, 
S_{\hbox{\tiny{gauge}}}(p_j,p_k) \ .
\ee
Constraints on this phase factor have been obtained from crossing symmetry in \cite{Janik}. 
The dressing factor can also be constrained through comparison with the leading quantum 
correction to the energy of semiclassical strings \cite{Sakura}-\cite{FK}. The quantum dressing 
factor \cite{HL} was in fact shown in \cite{AF} to satisfy the crossing equations, and 
a solution to the crossing relation has recently been proposed in \cite{Bcrossing}. 
  
In many occasions, underlying an integrable system there is a Hopf algebra of symmetries 
(see for instance \cite{qg} and references therein). Factorizable $S$-matrices 
can then always be written in the form 
\be
S_{12} = S_{12}^0 R_{12} \ ,
\label{Smatrix}
\ee
where $R_{12}$ is the intertwiner $R$-matrix for the Hopf algebra of symmetries, 
and the factor $S_{12}^0$ is the dressing phase. The magnons entering into the scattering 
process correspond to irreps $V_{\pi_i}$ of the Hopf algebra, and the $R$-matrix 
\be
R_{\pi_1 \pi_2}: V_{\pi_1} \otimes V_{\pi_2} \rightarrow V_{\pi_2} \otimes V_{\pi_1} 
\ee
is determined by the intertwiner condition
\be
R_{\pi_1 \pi_2} \Delta_{\pi_1 \pi_2} (a) = \Delta_{\pi_2 \pi_1} (a) R_{\pi_2 \: \pi_1} \ ,
\label{intert}
\ee
with $\Delta(a)$ the co-multiplication for an arbitrary element $a$ in the Hopf algebra. The 
physical meaning of the co-multiplication is to provide the composition law that defines the 
action of symmetry transformations on multimagnon states. From the intertwiner condition 
(\ref{intert}) the non-triviality of the $R$-matrix follows as a consequence of non-symmetric 
co-multiplications, {\em i.e.} ``non-classical'' composition laws. In the case of the planar 
limit of ${\cal N}=4$ Yang-Mills, the $S$-matrix is derived by imposing the intertwiner 
condition (\ref{intert}) with a non-symmetric co-multiplication for the generators of the 
$SU(2|2)$ algebra \cite{BeisertS}.
  
In case the underlying Hopf symmetry algebra contains a non-trivial central Hopf subalgebra 
\cite{DeConcini}, new interesting features appear. In particular, not all magnon irreps 
are allowed, and there is not a universal $R$-matrix such that
\be
R_{\pi_1 \pi_2} = \pi_1 \otimes \pi_2 R \ .
\label{R}
\ee 
In fact, independently of what the intertwiner $R$-matrix is, we should require, for any 
element $a$ in the central Hopf subalgebra, that
\be
\Delta_{12} (a) = \Delta_{21} (a) \ . 
\label{F}
\ee
This condition, with a non-symmetric co-multiplication, restricts the allowed irreps, that 
are now parameterized by the eigenvalues of the central elements, to live on certain Fermat 
curves in the spectrum of the central Hopf subalgebra. \footnote{These constraints on the 
allowed irreps have important implications for the existence of a universal $R$-matrix. In 
fact, $R_{\pi_1 \pi_2}$ as defined in (\ref{R}) should in principle exist for any couple of 
irreps, independently of whether they satisfy condition (\ref{F}).} In this note we will identify 
a central Hopf symmetry subalgebra ${\cal Z}$ for planar ${\cal N}=4$ Yang-Mills, with 
three generators and non-symmetric co-multiplications that will restrict the allowed 
magnon irreps to those with the BMN-like dispersion relation found in \cite{BeisertS} 
from supersymmetry. Multimagnon physical states are then defined as those invariant under 
the central Hopf subalgebra leading to the total zero momentum Virasoro condition. The Yang-Mills 
coupling enters through the undetermined constants parameterizing the Fermat curves that solve 
condition (\ref{F}). The central Hopf subalgebra governing the dispersion relation for magnons is 
generated by the two central elements introduced in \cite{BeisertS}, and an additional central 
generator $K$ related to the magnon momentum. This Hopf subalgebra is isomorphic to the 
central Hopf subalgebra of ${\cal U}_{q}(\widehat{SL(2)})$ with $q$ a root of unity. It is 
known that ${\cal U}_{q}(\widehat{SL(2)})$ is the affine Hopf symmetry algebra of the 
sine-Gordon model \cite{Bernard}, with $q$ determined by the sine-Gordon coupling. 
At the special value of $q$ a root of unity a central Hopf subalgebra, isomorphic to the one 
that we have identified for ${\cal N}=4$ Yang-Mills, is dynamically generated. In this case a 
new dispersion relation for the sine-Gordon solitons can be derived from the central elements 
in a way analogous to the one we have used in order to reconstruct the magnon dispersion 
relation. 
  
Finally, let us just briefly comment on the crossing transformation. It was originally 
suggested in \cite{Smirnov} that crossing for an affine Hopf algebra could be defined  
by promoting the action of the antipode into a certain change in the affine spectral 
parameter, the ``crossing transformation'', that becomes an inner automorphism of the 
algebra. This, together with the property of the universal $R$-matrix
\be
(\gamma \otimes \II) R = R^{-1} \ ,
\label{universal}
\ee
leads to a purely algebraic implementation of crossing symmetry. 
This program was developed for the sine-Gordon model in \cite{Bernard} by imposing invariance 
under the Drinfeld quantum double ${\cal D}({\cal A},{\cal A}^*)$ \cite{Drinfeld}, with 
${\cal A}$ the quantum affine Hopf algebra of the sine-Gordon model, and ${\cal A}^*$ the 
dual algebra. In \cite{Janik} this approach was suggested as a way to define crossing 
for ${\cal N}=4$ Yang-Mills. However, for ${\cal N}=4$, as well as for other 
integrable models enjoying invariance under a non-trivial central Hopf 
subalgebra, as for instance the chiral Potts model, an intrinsically algebraic 
definition of crossing transformations can be given independently of the assumption of 
existence of a universal $R$-matrix. The idea is simply to realize that the rapidity 
plane is the submanifold in the spectrum of the central subalgebra defined by 
the intertwiner conditions. Therefore, as a simple application of Schur's lemma 
we can lift to the rapidity plane the action of the antipode on the generators of the 
central subalgebra, and thus define crossing transformations to be this lifted action 
of the antipode $\gamma$ on the rapidity plane.
  
The article is organized as follows. In section 2 we will translate the dynamic 
$SU(2|3)$ chain into the existence of an abelian central Hopf subalgebra of symmetries, 
${\cal Z}$, with non-trivial co-multiplication rules and antipodes. We will show how the 
generators of this algebra turn out to be the central elements added to $SU(2|2)$ in order to 
induce the $SU(2|3)$ dynamics. We will then describe how the possible irreps, parameterized 
by the eigenvalues of the generators of the central algebra, are constrained from 
intertwiner conditions. These conditions determine the dispersion relation for 
magnons, and the elliptic curve on the rapidity plane. In section~3 we will wonder about 
the underlying Hopf algebra with ${\cal Z}$ as central subalgebra. We will provide evidence 
that it should correspond to some quantum Hopf affine algebra at a root of unity, with 
the central Hopf subalgebra ${\cal Z}$ as the enlarged center at such a root of unity. 
In section~4 we will identify the special features 
of the magnon kinematics with the conditions for the existence of a non-trivial 
center for the underlying quantum affine symmetry of the sine-Gordon model.


\section{The central Hopf subalgebra}

In this section we will describe the geometry underlying the Hopf algebra symmetry of 
the central extensions of the integrable $SU(2|2)$ chain. In particular, we will 
find the origin of the Virasoro constraints and the dispersion relation on purely 
algebraic grounds. But before doing that we will review in some detail the $SU(2|3)$ 
dynamic chain, and a convenient choice of representations. 


\subsection{Dynamics and representations}

The $S$-matrix of planar ${\cal N}=4$ supersymmetric Yang-Mills 
can be constructed using the fact that the 
complete $PSU(2,2|4)$ algebra splits into two equal pieces. Both $SU(2|2)$ factors 
share a central charge that behaves as the hamiltonian, which is part of the symmetry 
algebra. But in order to deal with the dynamical properties of the chain two extra 
central elements to the $SU(2|2)$ algebra need to be introduced \cite{BeisertS}. 
These additional central elements act trivially on physical states of vanishing total 
momentum. However they act in a non-trivial way on the magnon constituents of the physical 
states and therefore are relevant to fix the scattering $S$-matrix. \footnote{From a physical 
point of view the role of this center is very similar to the one played by the center 
$\mathbb{Z}_N$ of $SU(N)$ in QCD. Physical states are singlets with respect to the center, 
but the quark constituents transform non-trivially under $\mathbb{Z}_N$.} 

Before constructing suitable representations for the $SU(2|2)$ spin chain, let us first recall 
the $SU(2|3)$ integrable system \cite{su23}. In the $SU(2|3)$ sector we have $12$ supercharges, 
$Q^{i}_{\a}$ and $G^{i}_{\a}$, with $i=1,2,3$ and $\a=1,2$, three complex scalar fields and two 
spinors. The algebra is enlarged with a $U(1)$ subalgebra generated by the 
interacting hamiltonian. The full and the interacting hamiltonians, $H$ and $\delta H$, satisfy
\ba
&& [H,Q] = \frac {1}{2} Q \ , \quad [H,G] = - \frac {1}{2} G \ , \nonumber \\
&& [\delta H,Q]=0 \ , \quad [\delta H,G]=0 \ .
\ea
The dynamical nature of the $SU(2|3)$ chain arises because we can find states with the same 
quantum numbers and energy, but different length, which is defined in terms of the number of 
constituents. Fluctuations between these states make the length of the chain a dynamical 
variable. In particular, the allowed fluctuations, between $\phi^{[1}\phi^2\phi^{3]}$ and 
$\psi^{[1}\psi^{2]}$, play a fundamental role in the definition of the different irreps.
  
In order to construct the irreps, let us for instance consider two supercharges $Q^3_1$ and $Q^3_2$, 
acting on the field $\phi^3$. They transform $\phi^3$ into $\psi^1$ and $\psi^2$, respectively. 
We will now act with  $Q^3_1 Q^3_2$ on a formal state $|\phi^3\phi^3 \rangle$, and transform 
then the resulting state $| \psi^{[1}\psi^{2]} \rangle$ into $|\phi^{[1}\phi^2\phi^{3]} \rangle$ 
through a fluctuation. After these formal manipulations, we remove one $\phi^3$ field 
from the original and final states, and read from this the action $Q^3_1 | \psi^2 \rangle$ 
as $| \phi^{[1} \phi^{2]} \rangle$. 
  
Now, we will move to the $SU(2|2)$ sector, with only $8$ supercharges $Q^{a}_{\a}$ 
and $G^{a}_{\a}$, where now $a=1,2$ and $\a=1,2$. In this case we only have two scalar 
fields, and therefore there are no allowed fluctuations. However we can still rely 
formally on the dynamics of the $SU(2|3)$ chain to construct irreps. Consider for 
instance the two supercharges $Q_1^1Q_2^1$ acting on the scalar field $\phi^1$. Through 
the same argument as above we can consider $Q_1^1Q_2^1 | \phi^1 \phi^1 \rangle$, 
which leads to $| \psi^{[1} \psi^{2]} \rangle$. Using a fluctuation we transform 
now this state into  $|\phi^{[1}\phi^2 Z^] \rangle$, where the field $Z$ is playing 
the role of $\phi^3$. After this we remove $\phi^1$ from the first and last states 
to obtain $Q_1^1 | \psi^2 \rangle \simeq \ | \phi^2 Z \rangle$ or, in 
general \cite{BeisertS},
\be
Q_{\a}^{i} | \psi^{\b} \rangle \simeq \e_{\a\b} \e^{ij} | \phi^j Z \rangle \ .
\ee
Similar formal manipulations, with the spectator field $Z$, can be employed to define 
the irreps for the remaining set of supercharges, $G^{a}_{\a}$. We can act on $\psi^1$ 
with $G_1^1$ and $G_1^2$ to produce, respectively, $\phi^1$ and $\phi^2$. Thus, 
$G_1^1 G_1^2 | \psi^1 \psi^1 \rangle$ leads to $| \phi^{[1} \phi^{2]} \rangle$, and we 
can now define a fluctuation relating $| \phi^{[1} \phi^{2]} \rangle$ and 
$| \psi^{[1} \psi^2 Z^{-1]} \rangle $. Using this we get 
$G_1^1 | \phi^2 \rangle \simeq \ | \psi^2 Z^{-1} \rangle$ or, in general \cite{BeisertS},
\be
G_{\a}^{a} | \phi^{b} \rangle \simeq \e_{\a\b} \e^{ab} | \psi^{\b} Z^{-1} \rangle \ .
\ee
  
A direct consequence of these dynamic irreps for $SU(2|2)$ is the existence of 
central terms. In fact, we find that
\be
Q_{\a}^a Q_{\b}^b |\Psi\rangle \simeq \e_{\a\b} \e^{ab} | \Psi Z \rangle \ ,
\label{fluctu}
\ee 
for any generic state $|\Psi \rangle$. This action defines in a natural way 
a central term $B$ in $SU(2|2)$, because with respect to this algebra $|\Psi\rangle$ 
and $|\Psi Z \rangle$ are indeed the same state. However, as we will discuss in section 3, 
the co-multiplication of this central term, and also of the central element $R$ associated 
to the $G_{\a}^a$ supercharges, is asymmetric and non-trivial. We will use this observation 
to translate the $SU(2|3)$ dynamics into a deformed co-multiplication for a central Hopf 
subalgebra.


\subsection{Dynamical co-multiplication rules}

Let us now introduce the $SU(2|2)$ symmetry algebra. 
It is generated by two bosonic generators, $R^a_{\: b}$ and ${\cal L}^{\a}_{\: \b}$, 
together with the supersymmetry generators $Q^{\a}_{\: b}$ and $G^a_{\: \b}$ with central 
charge $c$,
\be
\{ Q^{\a}_a,G^b_{\b} \} = \delta_a^{\: b} {\cal L}^{\a}_{\: \b} + 
\delta^{\a}_{\: \b} R^b_{\: a} + \delta_a^{\: b} \delta_{\b}^{\: \a} c \ .
\ee
Following \cite{BeisertS}, we will extend the algebra with two central charges $B$ and $R$, 
\ba
\{ Q^{\a}_{\: a}, Q^{\b}_{\: b} \} = \e^{\a\b} \e_{ab} B \ , \nonumber \\
\{ G^{a}_{\: \a}, G^{b}_{\: \b} \} = \e^{ab} \e_{\a\b} R \ . 
\ea
The first thing to be noticed concerning these additional central elements is that they define 
an action on multiple magnon states with a non-trivial and asymmetric co-multiplication. In order 
to exhibit this co-multiplication, a new element needs to be introduced in the algebra 
through the relation
\be
K | \Psi \rangle = | Z \Psi \rangle \ ,
\label{K}
\ee
where $|\Psi\rangle$ denotes a generic magnon state. Following \cite{BeisertS}, an excitation 
with a given momentum $p$ will be
\be
|\Psi \rangle = \sum_n e^{ipn} \,
|\ldots Z \ldots {\Psi}_n \ldots Z \ldots \rangle \ .
\ee
Therefore, inserting or removing a field $Z$ on the excited state will correspond to
\be
|Z^{\pm} \Psi \rangle = \sum_n e^{ipn} | \ldots Z \ldots {\Psi}_{n\pm1} \ldots Z \ldots \rangle 
= e^{\mp ip} |\Psi \rangle \ ,
\ee
and thus we find
\be
K^{\pm 1} | \Psi \rangle = z^{\pm1} | \Psi \rangle \ ,
\ee
with $z$ the eigenvalue $z \equiv e^{-ip}$. It is immediate to check now that $K$ commutes 
with all the generators of $SU(2|2)$, and thus belongs to the center of the algebra. 
With this new operator we easily find the following co-multiplications for $B$, $R$ and $K$, 
\footnote{In all our equations we implicitly use a graded multiplication defined 
by 
\[
(a\otimes b)(c \otimes d) = (ac \otimes bd) (-1)^{[b][c]} \ .
\]
}
\ba
\Delta B & = & B \otimes K + \II \otimes B \ , \nonumber \\
\Delta R & = & R \otimes \II + K^{-1} \otimes R \ , \label{comul} \\
\Delta K & = & K \otimes K \ . \nonumber 
\ea
The operators $B$, $R$ and $K$ define with the co-multiplication (\ref{comul}) an abelian 
Hopf subalgebra, that we will denote by ${\cal Z}$. 


\subsection{$\hbox{Spec} \: {\cal Z}$ geometry and the dispersion relation}

We will now use invariance under the central Hopf subalgebra ${\cal Z}$ as a first step 
to relate integrability in the planar limit of ${\cal N}=4$ supersymmetric Yang-Mills with 
the existence of an underlying Hopf algebra symmetry. Let us first show how invariance 
under the central subalgebra already implies non-trivial constraints on the allowed intertwiners. 
We will assume, independently of what the underlying Hopf algebra governing the integrable 
structure of ${\cal N}=4$ Yang-Mills is, that the subalgebra ${\cal Z}$ is part of its center. 
We can now read from the co-multiplications (\ref{comul}) that the central Hopf subalgebra 
must be equipped with an antipode 
\ba
\gamma(B) & = & - B K^{-1} \ , \nonumber \\
\gamma(R) & = & - K R \ , \label{ant1} \\
\gamma(K) & = & K^{-1} \ . \nonumber
\ea
Notice that the antipode is non-trivial because of the non-trivial co-multiplication implied 
by the dynamics of the spin chain. Using now Schur's lemma we will characterize each irrep 
by the eigenvalues of the generators of ${\cal Z}$, 
\be
\pi(B) = x \ , \qq \pi(R) = y \ , \qq \pi(K) = z \ .
\label{Schur}
\ee
Next we will introduce a manifold $\hbox{Spec} \: {\cal Z}$ as the spectrum of ${\cal Z}$ 
\cite{DeConcini}, and use Schur's lemma to construct a map from the space of irreps into 
$\hbox{Spec} \: {\cal Z}$. Now, given two different irreps parameterized by $(x_1,y_1,z_1)$ 
and $(x_2,y_2,z_2)$, existence of an intertwiner requires
\be
\Delta_{12} (a) = \Delta_{21} (a) \ , 
\quad \forall \: a \in {\cal Z} \ .
\label{intertwiner}
\ee
When the co-multiplication is non-trivial this condition leads to relations between both 
irreps. Using now the co-multiplication (\ref{comul}) and the map into $\hbox{Spec} \: {\cal Z}$ 
defined by (\ref{Schur}), condition (\ref{intertwiner}) leads to the following set of curves 
of Fermat type in $\hbox{Spec} \: {\cal Z}$,
\be
\frac{x}{z-1} = \alpha, \quad \frac{y}{z^{-1}-1} = \beta \ ,
\label{Fermat}
\ee
with $\a$ and $\b$ some undetermined constants. Intertwiners will then only exist for 
irreps satisfying (\ref{Fermat}). In the notation of reference \cite{BeisertS}, we have 
$x=ab$ and $y=cd$. Notice that condition (\ref{Fermat}) are precisely those introduced 
in \cite{BeisertS} on the eigenvalues of the central elements $B$ and $R$, 
$ab=\a(e^{ip}-1)$ and $cd=\b(e^{-ip}-1)$, respectively. These relations arise in 
\cite{BeisertS} by imposing invariance under the central elements of multi-magnon 
physical states, with vanishing total momentum. However, these relations 
have a meaning of their own, independently of the condition of vanishing total 
momentum: they determine the explicit form of the single magnon dispersion relation. 
Let us also stress that in the previous derivation we have identified the origin of 
these relations directly from the structure of the central Hopf algebra and the 
intertwiner condition. The origin of (\ref{Fermat}) is thus independent of the condition 
of vanishing total momentum on physical states.~\footnote{Notice also that the 
interpretation of the central elements as gauge transformations, 
$B | \Psi \rangle = \a ( K |\Psi \rangle - |\Psi \rangle )$,  
$R | \Psi \rangle = \b ( K^{-1}|\Psi \rangle - |\Psi \rangle )$ 
is only valid once we have imposed the Virasoro constraints (\ref{Fermat}).} 
In fact, this condition simply means that physical states are singlets with respect to the 
central algebra ${\cal Z}$, in the same way as in QCD physical states are singlets under 
$\mathbb{Z}_N$. The dependence on the Yang-Mills coupling constant appears through the 
arbitrary constants $\alpha$ and $\beta$ characterizing the intertwiner Fermat curve. In order 
to recover the BMN scaling formula \cite{BMN} the choice $\a\b=2g^2$ needs to be 
done \cite{BeisertS}.

We can now use the intertwiner condition (\ref{Fermat}) together with the constraint imposed 
by the closure of $\{Q,G\}$, $ad-bc=1$, 
to solve for the central extension,
\be
c(z) = \pm \frac {1}{2} \sqrt{1+4\a\b(2-z-z^{-1})} \ .
\label{c(z)}
\ee
The region in $\hbox{Spec} \: {\cal Z}$ on which intertwiners for arbitrary pairs of points 
exist is thus the branch cover of the $z$-plane defined by the function $c(z)$. In fact irreps 
for which an intertwiner exist are characterized by the pair $(z,\pm c(z))$. The 
plus sign will correspond to irreps for particles, and the minus sign to antiparticle irreps.


\subsection{The kinetic plane}

We will now parameterize different irreps using $z$ and $c(z)$ as coordinates, 
and we will identify the two possible branches of $c(z)$ with particle and antiparticle 
irreps. Let us first translate these coordinates into the ones employed in \cite{BeisertS}, 
$x^{\pm}$. We get 
\be
x^-(z) = \frac {i}{2g} \frac {1 \pm 2c(z)}{(z-1)} \ ,
\label{x(z)}
\ee
while $x^+ = z x^-(z)$, which together with $c(z)$ define a double covering of the $z$-plane. 
At the self-dual point $z=1$ with respect to the antipode transformation, $z \rightarrow 1/z$, 
$x^{\pm}$ goes to infinity for the positive branch of $c(z)$, or to zero for the negative 
branch. The magnon charges are now given by
\be
q_r(z) = \frac {i(-2ig)^{r-1}}{r-1} \left( \frac {z-1}{1\pm 2c(z)} \right)^{r-1} 
\left( \frac {1}{z^{r-1}} -1 \right) \ ,
\label{charges}
\ee
which provides two different values, $q_r^{\pm}$, depending on the choice of branch for 
$c(z)$. They correspond to the magnon charges for particles and antiparticles, respectively. 
Moving from one particle irrep into an antiparticle irrep amounts to a change in $z$ 
along a path going through the branch cuts of $c(z)$. In particular, the branch points 
of $c(z)$ are located at
\be
z^{\pm} = \frac {1}{8\a\b} \big[ 1 + 8 \a\b \pm \sqrt{1+16\a\b} \, \big] \ .
\label{branch}
\ee 
Together with $z=0$, the branch points $z=z^{\pm}$ define an elliptic curve. When written in 
Weierstrass form,
\be
y^2 = 4x^3 - g_2 x - g_3 \ ,
\ee
the elliptic invariants are 
\ba
g_2 & = & \frac {1}{12} (1 + 16\a\b + 16 \a^2\b^2) \ , \nonumber \\
g_3 & = & \frac {1}{216} (1+8\a\b) (-1-16\a\b + 8\a^2\b^2) \ .
\ea
This is precisely the curve derived in \cite{Janik} in order to implement crossing on a 
generalized rapidity plane. In the strong coupling regime $\a\b \rightarrow \infty$ the branch 
points $z^{\pm} \rightarrow 1$, and the curve degenerates to 
\be
y^2=z(z-1)^2 \ .
\ee
\begin{figure}[t]
\begin{center}
\includegraphics[height= 4 cm,angle= 0]{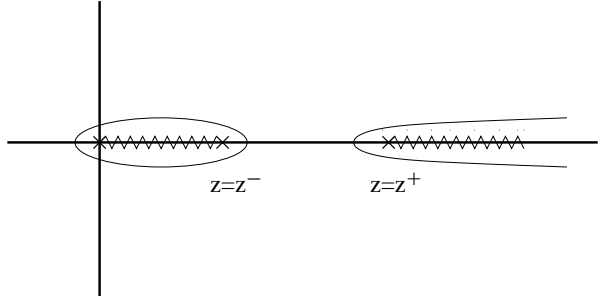}
\end{center}
\caption{The algebraic curve $y^2=z(z-z^-)(z-z^+)$.}
\end{figure}
  
Let us now discuss the different magnon irreps in the strong and 
weak coupling regimes. We will use $x$, $y$ and $z$, subject to the 
constraint (\ref{Fermat}), to parameterize the diverse irreps. And we will employ $z$ 
as the fundamental parameter, without assuming $e^{ip}$ as a particular representation. 
We first consider the strong coupling regime defined by $\a, \b \gg1$. When both 
$\a$ and $\b$ are large there are two possible irreps: those with generic $x$ and $y$ 
eigenvalues, but with $z$ close to $1$, and those with generic values of $z$, while 
$x$ and $y$ are taken to be large. We will refer to these representations as of type 
I and type II, respectively. The magnon energy for irreps of type I is
\be
c^{I}(z) = \pm \frac {1}{2} \sqrt{1+xy} \ ,
\label{I}
\ee
which is finite for both particle and antiparticle irreps. If we now read $z$ as 
$e^{ip}$, type I irreps are those with momentum $p$ close to zero. In the case of irreps of 
type II the momentum $p$ can reach generic values, but the magnon energy is 
\be
c^{II}(z) \simeq \pm \sqrt{xy} \ .
\label{II}
\ee
The existence of these two kinds of irreps is crucial in order to understand the 
strong coupling regime of semiclassical strings. As we reach the strong coupling regime 
the genuine BMN irreps, with generic and finite values of $x$ and $y$, and 
$z \simeq 1$, will start competing with those with generic $z$, but large values 
of $x$ and $y$. However, those irreps with arbitrary values of $z$ will be decoupled 
in the strong coupling limit, because its energy will diverge, while the 
ones with $z \simeq 1$ have finite energies. Thus the contribution of 
irreps of type I is the most natural one to the strong coupling region. Notice 
also that in this approach the coupling constant is a free parameter labeling 
different Fermat curves in $\hbox{Spec} \: {\cal Z}$. Thus, the condition of 
strong coupling simply fixes a certain curve with large values of $\a$ and~$\b$. 
All points on this curve are natural contributions to the strong coupling 
limit of the $S$-matrix. 
  
We will now analyze the weak coupling regime, where both $\a$ and $\b$ are close 
to zero. In this region, irreps satisfying the intertwiner condition 
(\ref{Fermat}) are those with generic values of $z$, but with $x$ and $y$ 
close to zero. For all of these irreps the energy is fixed at $c(z)= \pm 1/2$. 
Therefore, the magnon charges (\ref{charges}) for the antiparticle irreps will diverge, 
something that we can interpret in terms of decoupling of antiparticles 
from the physical spectrum. In the weak coupling region we also find 
irreps with $z$ close to one, but with $x$ and $y$ both equal or close 
to zero. 
 

\subsection{The $SU(1|2)$ $S$-matrix and crossing transformations}
  
As it was shown in \cite{BeisertS}, the $SU(2|2)$ $S$-matrix can be directly derived 
from the $S$-matrix for the $SU(1|2)$ sector. In this section we will briefly discuss 
the derivation of the $SU(1|2)$ $S$-matrix, and the meaning of the crossing 
transformations. Let us consider two irreps 
$V_{\pi_1}$ and $V_{\pi_2}$ of $SU(1|2)$, and let us define the $R$-matrix as 
\be
R(1,2) = \sum_i S^{i}(1,2) P^{i}_{1,2} \ , 
\label{r12}
\ee
where $P^{i}_{1,2}$ are the projector intertwiners, and with the sum extending over the 
$V_{\pi_i}$ irreps in the decomposition of $V_{\pi_1} \otimes V_{\pi_2}$. In order to fix the 
functions $S^{i}(1,2)$ we will impose the intertwiner condition
\be
S(1,2) \Delta_{\pi_1\pi_2}(a) = \Delta_{\pi_2\pi_1}(a) S(1,2) \ ,
\label{int2}
\ee
for any element in the $SU(1|2)$ algebra, and where $S(1,2) = P R(1,2)$. The solution to (\ref{int2}) 
if we consider symmetric co-multiplications, $\Delta(a) = a \otimes \II + \II \otimes a$, is the 
trivial one $R(1,2) = S^{0}(1,2) \sum _i P^{i}_{1,2} = S^0(1,2) \II$. Since we are interested 
in the $SU(2|2)$ $S$-matrix, we can try to fix the functions $S^{i}(1,2)$ in (\ref{r12}) by 
imposing the intertwiner condition (\ref{int2}), but for those elements on $SU(2|2)$ 
that are not in $SU(1|2)$. Denoting by $\tilde{Q}$ and $\tilde{G}$ the additional 
supersymmetry generators, we can recover the $SU(2|2)$ $S$-matrix by imposing (\ref{int2}) 
with 
\ba
\Delta(\tilde{Q}) & = & \tilde{Q} \otimes K + \II \otimes \tilde{Q} \ , \label{cosu12} \nonumber \\
\Delta(\tilde{G}) & = & \tilde{G} \otimes K^{-1} + \II \otimes \tilde{G} \ ,
\ea
where the generator $K$ is the one introduced in (\ref{K}).
  
It is worth to compare this construction with the one in \cite{Delius}. In this reference the 
affinization of the $R$-matrix for a quantum deformation of the $SU(1|2)$ group was considered. 
The $R$-matrix was constrained by two conditions, which are the intertwiner condition (\ref{int2}) 
for the generators of $SU(1|2)$ with a deformed co-multiplication, and an additional 
intertwiner condition involving the extra generators of the affine algebra. This new 
condition depends on the spectral parameter, which turns to be the rapidity. The main difference 
with the previous construction is that the additional intertwiner conditions in the 
case of $SU(2|2)$ are not associated with any form of affinization. Moreover, the magnon 
rapidities enter into the $R$-matrix through the co-multiplication (\ref{cosu12}) leading 
to an $S$-matrix depending on the two magnon rapidities. 
  
Let us now consider the issue of crossing. As we have already discussed in the previous subsections, 
due to the existence of a central Hopf subalgebra ${\cal Z}$ the different magnon irreps 
can be characterized by the eigenvalues of the central elements. In this case, we can lift the 
action of the antipode on the generators of ${\cal Z}$ to the space of irreps. We thus define 
\be
\bar{\pi}(a) = \pi \big( \gamma (a) \big) \ , \quad \forall a \in {\cal Z} \ . 
\ee
This map leads to
\ba
x & \rightarrow & \bar{x} = -z^{-1} x \ ,\nonumber \\
y & \rightarrow & \bar{y} = -zy \ , \\
z & \rightarrow & \bar{z} = z^{-1} \ . \nonumber 
\ea
Notice now that the identification of $\hbox{Spec} \: {\cal Z}$ with the rapidity plane allows a 
kinematical definition of crossing as transformation on the rapidity plane which is completely 
determined by the definition of the antipode for the generators of ${\cal Z}$. In this 
sense, this implementation of crossing only depend on the central Hopf subalgebra. This 
definition of crossing on the space of irreps is completely independent of the existence 
of a universal $R$-matrix. We can represent the previous implementation of crossing 
through
\be
\begin{CD}
{\cal Z} @>\pi>> \hbox{Spec} \: {\cal Z} \\
@V{\gamma}VV   @VV{\hbox{\footnotesize{cross}}}V \\
{\cal Z} @>\pi>> \hbox{Spec} \: {\cal Z}
\end{CD} 
\ee
where $\pi$ is the map from the central subalgebra into the rapidity manifold 
given by Schur's lemma, $\gamma$ is the antipode in ${\cal Z}$ and ``cross'' is the 
crossing transformation. Notice that cross preserves the curves obtained 
from the intertwiner condition. 


\section{The Hopf algebra symmetry}

In the previous section we have identified an abelian Hopf subalgebra ${\cal Z}$ that 
encodes the information on the multi-magnon states of vanishing total momentum through 
the intertwiner conditions. In order to construct the Hopf subalgebra ${\cal Z}$ 
we have employed the central elements of $SU(2|2) \ltimes \mathbb{R}^2$, together 
with the generator $K$. In this section we will pose the question of the existence 
of a Hopf algebra ${\cal A}$ with central extension ${\cal Z}$, in such a way that 
points in $\hbox{Spec} \: {\cal Z}$ are in a one-to-one correspondence with irreps of ${\cal A}$.
  
The most natural candidate to ${\cal A}$, which cannot a priori be identified with 
$SU(2|2) \ltimes \mathbb{R}^2$, is a quantum group Hopf algebra, with the quantum deformation 
parameter at a particular root of unity, because quantum groups at roots of unity exhibit 
an enlarged central Hopf subalgebra. Let us consider to clarify ideas a Hopf algebra 
with generators $E_i$, $F_i$ and $K_i$ in the Cartan-Chevalley basis, and with $q^l=1$. 
In this case $E_i^l$, $F_i^l$ and $K_i^l$ are part of the generators of the central 
Hopf subalgebra. It is important to stress that these central elements are not added 
to the algebra ${\cal A}$, but rather arise as a consequence of the quantum deformation 
parameter $q$ being a root of unity. With this observation in mind, what we are searching 
for must be a Hopf algebra and a particular value of $q^l=1$, such that the corresponding 
central subalgebra is isomorphic to the one defined through (\ref{comul}) in the 
${\cal N}=4$ Yang-Mills case. This would provide a natural explanation on the origin 
of the additional central elements extending the $SU(2|2)$ algebra. 
  

\subsection{Cyclic two-dimensional irreps}

In order to uncover the algebra ${\cal A}$, let us first consider the kind of periodic 
irreps that we can define in $SU(2|2)$. We can construct two-dimensional subspaces by 
the following cycle of transformations,
\be
\ldots \overset{Q_{\b}^{j}}{\longrightarrow}
| \phi^{i} \rangle \overset{Q_{\a}^{i}}{\longrightarrow} | \psi^{\a} \rangle 
\overset{Q_{\b}^{j}}{\longrightarrow} |\phi^{i} \rangle 
\overset{Q_{\a}^{i}}{\longrightarrow} \ldots \ ,
\ee
with $i \neq j$ and $\a \neq \b$. In a similar way we can define 
\be
\ldots \overset{G_{\b}^{j}}{\longrightarrow}
| \psi^{\a} \rangle \overset{G_{\a}^{i}}{\longrightarrow} | \phi^{i} \rangle 
\overset{G_{\b}^{j}}{\longrightarrow} |\psi^{\a} \rangle 
\overset{G_{\a}^{i}}{\longrightarrow} \ldots 
\ee
It is of course very tempting, once we have these two-dimensional vector spaces, 
to interpret them as some sort of cyclic irreps. In particular, we will identify them 
with the two-dimensional cyclic irreps of ${\cal U}_{q}(\widehat{SL(2)})$ with $q^4=1$. 
In fact, we can define operators
\ba
\tilde{Q}^{ij}_{\a\b} & \equiv & Q^{i}_{\a} + Q^{j}_{\b} \ , \nonumber \\
\tilde{G}^{ij}_{\a\b} & \equiv & G^{i}_{\a} + G^{j}_{\b} \ ,
\ea 
satisfying
\ba
(\tilde{Q}^{ij}_{\a\b})^2 & = & \e^{ij} \e_{\a\b} B \ , \label{tildeQG} \nonumber \\
(\tilde{G}^{ij}_{\a\b})^2 & = & \e^{ij} \e_{\a\b} R \ .
\ea
Then, the cyclic representation can be chosen as
\ba
&& 
\ldots \overset{\tilde{Q}_{21}^{12}}{\longrightarrow} 
| \phi^{1} \rangle \overset{\tilde{Q}_{21}^{12}}{\longrightarrow} | \psi^{2} \rangle 
\overset{\tilde{Q}_{21}^{12}}{\longrightarrow} |\phi^{1} \rangle 
\overset{\tilde{Q}_{21}^{12}}{\longrightarrow} \ldots \ , \label{cyclic} \nonumber \\
&& 
\ldots \overset{\tilde{G}_{21}^{12}}{\longrightarrow} 
| \psi^{2} \rangle \overset{\tilde{G}_{21}^{12}}{\longrightarrow} | \phi^{1} \rangle 
\overset{\tilde{G}_{21}^{12}}{\longrightarrow} |\psi^{2} \rangle
\overset{\tilde{G}_{21}^{12}}{\longrightarrow} \ldots \ .
\ea
In this way we can think of (\ref{cyclic}) as a cyclic irrep of 
${\cal U}_{q}(\widehat{SL(2)})$ with $q^4=1$, where $\tilde{Q}_{21}^{12} \sim E$ and 
$G_{21}^{12} \sim F$, and with $B$ and $R$ being parts of the central Hopf subalgebra 
of ${\cal U}_{q}(\widehat{SL(2)})$ at $q^4=1$. This should be just considered as 
a formal hint toward the challenge of uncovering the underlying Hopf symmetry algebra ${\cal A}$ 
whose central subalgebra ${\cal Z}$ is the central Hopf subalgebra of ${\cal N}=4$ 
Yang-Mills. From (\ref{tildeQG}) it follows that $\tilde{Q}^{ij}_{\a\b} = \tilde{Q}^{ji}_{\b\a}$, 
so that there are only two different $\tilde{Q}$ operators, $\tilde{Q}^{12}_{21}$ and 
$\tilde{Q}^{12}_{12}$. The same holds true for the $\tilde{G}$ operators. Thus, 
we find the adequate number of operators for the map to the affine 
${\cal U}_{q}(\widehat{SL(2)})$ at $q^4=1$, which therefore appears 
as a natural candidate to at least part of the underlying Hopf algebra 
of the planar limit of ${\cal N}=4$ Yang-Mills. 

As a step further, we will also suggest a co-multiplication for the supersymmetry generators 
$Q$ and $G$, consistent with the one defining ${\cal Z}$. Dynamics was introduced in 
representation theory through fluctuations of the form (\ref{fluctu}). A suitable formal 
way to respect these fluctuation equivalences would be the definition of irreps of the type 
$Q_{\b}^{i} \phi^{i} \sim \psi^{\b} Z^{1/2}$, and $Q_{\a}^{i} \psi^{\b} \sim \phi^{i} Z^{1/2}$. 
The advantage of these definitions of irreps is formal consistency with the fact 
that the central elements $B$ and $R$ have non-trivial co-multiplications and, 
as consequence, non-trivial antipodes. Thus we will formally extend the 
Hopf algebra structure by requiring 
\ba
\Delta Q & = & Q \otimes \II + {\cal K} \otimes Q \ , \nonumber \\
\Delta G & = & G \otimes \II + {\cal K}^{-1} \otimes G \ , 
\label{co}
\ea 
where ${\cal K}$ would be part of the Cartan subalgebra of ${\cal A}$, and 
such that  ${\cal K}^2=K$, with $K$ the generator in ${\cal Z}$. 
The corresponding antipodes are
\ba
\g(Q) & = & - {\cal K}^{-1} Q \ , \nonumber \\
\g(G) & = & - {\cal K} G \ . 
\label{ant}
\ea
Notice that now (\ref{co}) and (\ref{ant}) are perfectly consistent with the co-multiplication 
and the antipode of ${\cal Z}$, as well as with relations (\ref{tildeQG}). 


\section{Magnon kinematics and the sine-Gordon model}

In this section we will explore the kinematics of giant magnons on semiclassical strings. 
Semiclassical strings moving in $\mathbb{R} \times S^2$ are equivalent to 
the sine-Gordon integrable model \cite{Pohlmeyer,Mikhailov}. The Virasoro constraints 
lead to 
\be
\big[ \partial_{\tau}^2 - \partial_{\s}^2 \big] \phi = - \frac {1}{2} \sin  (2\phi) \ .
\ee
This sine-Gordon model corresponds to a particular value of the coupling constant, 
$\b=2$.~\footnote{We are normalizing the sine-Gordon model as 
\[
S = \frac {1}{4\pi} \int d^2z \partial_z \phi \partial_{\bar{z}} \phi + 
\frac {\lambda}{\pi} \int d^2z \cos (\b \phi) \ .
\]
Thus, in the string case we have $\b=2$ and $\l=\frac {1}{8}$ ({\em i.e.} $m^2=1$).}

This giant magnons of \cite{magnons} correspond precisely to sine-Gordon 
solitons, with the only difference that the magnon energy goes like the inverse 
of the soliton energy. Introducing the sine-Gordon rapidity $\th$, the 
soliton energy is given by
\be
E_{\hbox{\tiny{sG}}} = \cosh \th \ . 
\ee
Through the map of the string sigma model into the sine-Gordon system we have
\be
\sin \left( \frac {p}{2} \right) = \frac {1}{\cosh \th} \ .
\ee
Thus, the magnon energy, given by the large coupling limit of the dispersion 
relation, $E \simeq \sqrt{\l}/\pi \sin (p/2)$, goes like 
$1/E_{\hbox{\tiny{sG}}}$. Notice also that the magnon energy $E \simeq 1/ \cosh \theta$ 
is strictly the same as the one of elementary excitations of the 
anti-ferromagnetic isotropic Heisenberg chain in the thermodynamic limit. In fact in this case 
we have $E(\theta) \simeq 1/\cosh \theta$ and
\be 
p(\theta) \simeq \pi -\tan^{-1} ( \sinh \theta ) \ ,
\ee 
which leads to the dispersion relation $E(p) \simeq \sin (p/2)$. Let us recall that the low lying 
excitations for the anti-ferromagnetic chain are the holes on the Dirac sea of Bethe strings, 
and they have spin $\frac{1}{2}$ \cite{FT}. The formal relation with the sine-Gordon model 
is hidden in the special form of the rapidity dependence of the momentum excitations.
  
If we move now into weaker coupling and interpret $E(p) =\sqrt{1+\lambda/\pi^{2} 
\sin^{2} (p/2) }$ as the relativistic relation $E^{2} = m^{2} +p^{2}$, 
we should identify $\lambda/\pi^2 \sin^2 (p/2)$ with 
the momentum square. Thus, a natural definition of rapidity is \cite{magnons} 
\be
\sinh^2 \th_p = \frac {\l}{\pi^2} \sin^2 \left( \frac {p}{2} \right) \ . 
\ee
From this identification we get
\be
e^{\theta_{p}} = \mp 2\sqrt{xy} \pm \sqrt{4xy+1}
\ee
Notice that in the strong coupling limit and for irreps of type I we have a generic value 
of the rapidity $\theta_{p}$.
  
Let us now try to understand the physical meaning of this rapidity in terms 
of the quantum symmetries of the sine-Gordon model. It is a well known result 
that the non-local charges in the sine-Gordon model generate an affine quantum 
algebra ${\cal U}_{q}(\widehat{SL(2)})$, with
\be
q = e^{- \frac {2 \pi i}{\b^2}} \ ,
\ee
so that $q^4=1$ when $\b=2$. This affine quantum algebra at $q^4=1$ is isomorphic 
to the ${\cal N}=2$ supersymmetry algebra, with generators $Q_{\pm}$ and $\bar{Q}_{\pm}$. We can now 
represent the generators of ${\cal U}_{q}(\widehat{SL(2)})$ in terms of the ones of 
${\cal U}_{q}({SL(2)})$, $E$, $F$ and ${\cal K}$, if we introduce an affine 
parameter, that plays the role of a rapidity, 
\ba
&& Q_+ = e^{\th} E  \ , \quad Q_- = e^{\th} F \ , \nonumber \\
&& \bar{Q}_+ = e^{-\th} F{\cal K} \ , \quad \bar{Q}_- = {\cal K}^{-1}e^{-\th} E \ . 
\label{affine}
\ea
For regular solitonic irreps we get from here the standard relation $p = \sinh \th$, 
with $p$ defined in terms of $Q_{\pm}$ and $\bar{Q}_{\pm}$ by the Serre relations. These 
are the standard sine-Gordon solitons that are directly connected at strong coupling 
with the giant magnons, up to the change in the energy relations. However 
for $q^4=1$ we also have classical irreps, with $Q_{\pm}$ and $\bar{Q}_{\pm}$ 
being non-vanishing elements in the center. For these non-classical irreps we also 
have $K ={\cal K}^2$ in the center of ${\cal U}_{q}(\widehat{SL(2)})$, together with 
$E^2$ and $F^2$. In this case we can use the elements in the center to construct a 
candidate for the momentum through \cite{Vafa} 
\be
P = Q_{\pm}^{2} \ , \quad \bar{P} = \bar{Q}_{\pm}^{2}
\ee
with the physical momentum $p=P- \bar{P}$. 
If we write $p^{2} = ( Q_{+}^{2} - \bar{Q}_{+}^{2})( Q_{-}^{2} - \bar{Q}_{-}^{2})$ 
we get, when $\theta =0$,
\be
p^{2} = \frac{\lambda}{\pi^{2}} (z-1)(z^{-1}-1) \ ,
\ee
for the eigenvalues of $E^{2}$, $F^{2}$ and $K={\cal K}^{2}$ given by 
$\sqrt{\lambda}(z-1)/2\pi$, $\sqrt{\lambda}(z^{-1}-1)/2\pi$ and $z$, 
respectively. This is just the relation $p^{2} = \lambda/\pi^{2} \sin^{2} (p/2)$ 
for $z=e^{-ip}$. This provides further evidence that magnons must be related to sine-Gordon 
solitons in non-classical irreps where the central subalgebra is realized in a non-trivial way. 
The kinematic arena for these magnons is determined by the Spec of the central 
subalgebra of symmetries of ${\cal U}_{q}(\widehat{SL(2)})$~\footnote{A similar
phenomena takes place in the chiral Potts model \cite{GS2}.}. Nicely enough, this central 
subalgebra on the string theory side is isomorphic to the central subalgebra ${\cal Z}$ for 
${\cal N}=4$ Yang-Mills. Thus, the constituents on both sides of the correspondence 
share the same kinematic arena, defined by the same central subalgebra. The study of 
these common features clearly deserves further analysis. 


\vspace{5mm}
\centerline{\bf Acknowledgments}

It is a pleasure to thank E. L\'opez for collaboration along different 
stages of this work and illuminating discussions. We are also indebted 
to L. \'Alvarez-Gaum\'e for continuous interest as this work evolved. 
R.~H. would also like to thank the organizers of the Albert Einstein 
Institut workshop on Integrability in Gauge and String Theory for hospitality and a 
stimulating atmosphere while this work was being completed. 
The work of C.~G. is partially supported by the Spanish DGI contract FPA2003-02877 
and CAM project HEPHACOS P-ESP-00346. 


\end{document}